\documentclass[aps,prb,twocolumn,superscriptaddress]{revtex4-2}
\usepackage{geometry}     
\usepackage{physics}    
\usepackage{lipsum}       
\usepackage{graphicx}
\usepackage{amssymb}
\usepackage{braket}
\usepackage{mathrsfs}
\usepackage{amsmath}
\usepackage{floatrow}
\usepackage{subfig}
\usepackage{makecell}
\usepackage[T1]{fontenc}
\usepackage[ruled]{algorithm2e}

\begin{document}

\title{Characterizing non-Markovian Quantum Process by Fast Bayesian Tomography}
\author{R. Y. Su}
\email{rocky.su@unsw.edu.au}
\affiliation{School of Electrical Engineering and Telecommunications, The University of New South Wales, Sydney, NSW 2052, Australia}
\author{J. Y. Huang}
\affiliation{School of Electrical Engineering and Telecommunications, The University of New South Wales, Sydney, NSW 2052, Australia}
\author{N. Dumoulin. Stuyck}
\affiliation{School of Electrical Engineering and Telecommunications, The University of New South Wales, Sydney, NSW 2052, Australia}
\affiliation{Diraq, Sydney, NSW, Australia}
\author{M. K. Feng}
\affiliation{School of Electrical Engineering and Telecommunications, The University of New South Wales, Sydney, NSW 2052, Australia}

\author{W. Gilbert}
\affiliation{School of Electrical Engineering and Telecommunications, The University of New South Wales, Sydney, NSW 2052, Australia}
\affiliation{Diraq, Sydney, NSW, Australia}
\author{T. J. Evans}
\affiliation{Centre for Engineered Quantum Systems, School of Physics, The University of Sydney, Sydney 2006, Australia}
\author{W. H. Lim}
\author{F. E. Hudson}
\affiliation{School of Electrical Engineering and Telecommunications, The University of New South Wales, Sydney, NSW 2052, Australia}
\affiliation{Diraq, Sydney, NSW, Australia}
\author{K. W. Chan}
\affiliation{School of Electrical Engineering and Telecommunications, The University of New South Wales, Sydney, NSW 2052, Australia}
\affiliation{Diraq, Sydney, NSW, Australia}
\author{W. Huang}
\altaffiliation[]{current address: ETH Zurich}
\affiliation{School of Electrical Engineering and Telecommunications, The University of New South Wales, Sydney, NSW 2052, Australia}
\author{Kohei M. Itoh}
\affiliation{School of Fundamental Science and Technology, Keio University, Yokohama, Japan}
\author{R. Harper}
\author{S. D. Bartlett}
\affiliation{Centre for Engineered Quantum Systems, School of Physics, The University of Sydney, Sydney 2006, Australia}
\author{C. H. Yang}
\author{A. Laucht}
\author{A. Saraiva}
\author{A. S. Dzurak}
\author{T. Tanttu}
\email{t.tanttu@unsw.edu.au}
\affiliation{School of Electrical Engineering and Telecommunications, The University of New South Wales, Sydney, NSW 2052, Australia}
\affiliation{Diraq, Sydney, NSW, Australia}

\date{\today}

\begin{abstract}
\par To push gate performance to levels beyond the thresholds for quantum error correction, it is important to characterize the error sources occurring on quantum gates. However, the characterization of non-Markovian error poses a challenge to current quantum process tomography techniques. Fast Bayesian Tomography (FBT) is a self-consistent gate set tomography protocol that can be bootstrapped from earlier characterization knowledge and be updated in real-time with arbitrary gate sequences. Here we demonstrate how FBT allows for the characterization of key non-Markovian error processes. We introduce two experimental protocols for FBT to diagnose the non-Markovian behavior of two-qubit systems on silicon quantum dots. To increase the efficiency and scalability of the experiment-analysis loop, we develop an online FBT software stack. To reduce experiment cost and analysis time, we also introduce a native readout method and warm boot strategy. Our results demonstrate that FBT is a useful tool for probing non-Markovian errors that can be detrimental to the ultimate realization of fault-tolerant operation on quantum computing.
\end{abstract}
\maketitle

\section{Introduction}
 
 \par The development of fault-tolerant and scalable quantum computers relies on achieving quantum gate fidelities that are beyond the error correction thresholds \cite{QEC, surfcode, linearlogicqubit}. However, qubits, the fundamental building blocks of quantum computers, reside in a noisy environment and are manipulated with imperfect control sequences.
 
 \par Over the past few decades, a range of quantum characterization, verification, and validation (QCVV) techniques have been developed to diagnose errors in quantum circuits. Randomized benchmarking (RBM) \cite{RB1, RB2, RB3}, a widely accepted metric in the community, is experimentally simple and efficient in characterizing gate performance. Although fidelity will indicate the overall performance of the gates, it does not provide information about the types of errors that degrade the gates.

\par Fully diagnosing the errors quantitively with quantum tomography protocols would help further mitigate errors in the gate set. Quantum process tomography (QPT) \cite{QPT} reconstructs a single gate process by applying an informationally complete ensemble of state preparation and measurement (SPAM) before and after the gate operation. However, it is not a calibration-free protocol, and yields biased results when SPAM are noisy. Self-consistent methods for gate set tomography \cite{selfconsistTomo, GST0, GST1} overcome this problem by reconstructing the whole gate set, including the noisy SPAM channels, so that no prior calibrations are required. 

\par Fast Bayesian tomography (FBT) \cite{timFBT} is a self-consistent gate set process tomography method which harnesses the power of Bayesian inference. It is a flexible and agile tool that allows arbitrary random sequences to be fed to the model, and FBT can be bootstrapped from the knowledge obtained from earlier characterization. Since the model is updated iteratively, sequence by sequence, the experiment and analysis can run simultaneously in realtime. 

\par In this paper, we use spin qubits to demonstrate that FBT can be used to probe and characterize non-Markovian errors in spin qubits. Noticeably, spatial-temporally correlated noise is prevalent on silicon quantum devices \cite{yoneda2022noise, BoterSiSiGeNoise}, being one of the major sources of non-Markovian behavior in our experiment. Such non-Markovian errors can impede the performance of quantum error correction \cite{nonMarkovianErrorCorr1, nonMarkovianErrorCorr2} and challenge most QCVV methods. Ideally, the performance of a gate is independent of its position in a quantum circuit and the lab time being performed. In other words, the gate processes are expected to be memory-less Markovian processes, which is a fundamental premise for quantum gate process tomography protocols. Methods to probe the context-dependent quantum errors have been proposed \cite{nonMarkovCharacterisation2, nonMarkovCharacterisation3, nonMarkovRB}, but it remains challenging at the quantum circuit level with gate set process tomography. A promising way to study non-Markovian dynamics for gate processes is via process tensor tomography (PTT) \cite{nonMarkovianPTT1, nonMarkovianPTT2}, which reconstructs the process tensor to obtain the spatial-temporal correlation of the gate processes in a sequence context. However, PTT is experimentally resource intensive and tricky to scale up with quantum systems. 

\par As with other protocols, without special treatment, non-Markovian errors degrade the performance of FBT. In addition, it has been unclear how FBT could identify the non-Markovian nature of the quantum process. We use a device consisting of a pair of spin qubits as a test bed. From the early stages of device characterization, we have observed that gate performance drifts slowly in hours-long experiments, in addition, performance appears to have sequence length dependency. In this paper, we propose two experimental designs for FBT to characterize the non-Markovian behavior of the gate processes. Based on the time scale where the noise is effective, we study the behavior in intra-sequence and inter-sequence regimes individually. The intra-sequence regime explores variations in the gate error process across different gate sequence lengths. The inter-sequence regime tracks the time evolution of the gate noise process by correlating intermediate FBT results with lab time. To minimize experimental costs for timing-sensitive experiments, we study the validity of using native measurement results (here, parity readouts) as the input for FBT. Moreover, we discuss several bootstrapping strategies for the initial priors of the gate set, which is critical to lower the experimental cost. To make gauge-variant metrics more consistent, we eliminate gauge ambiguity by implementing gauge optimization for FBT.

\section{Results}
\subsection{Online FBT setup}

 \begin{figure*}[tb]
   \centering
   \includegraphics[width=\columnwidth]{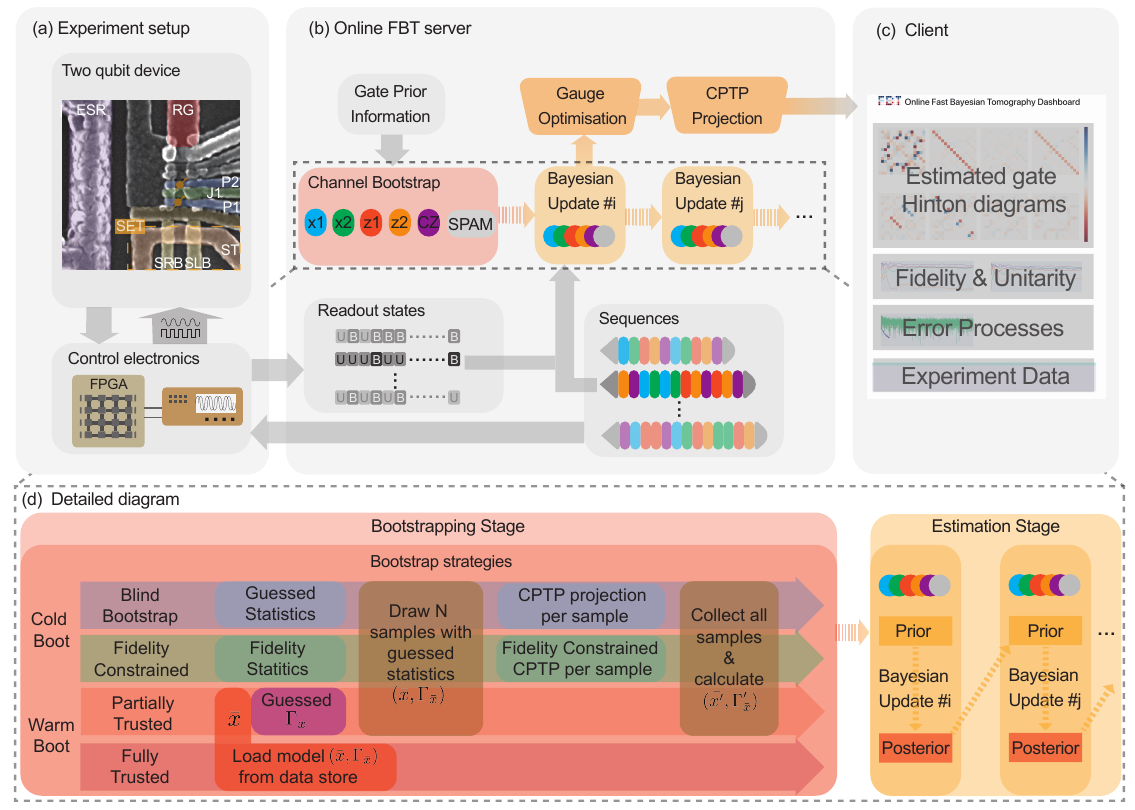}%
   
   \caption{High-level schematic of web-based online FBT service. (a) On the top is the scanning electron micrograph (SEM) image of a device similar to the one used in the experiment. A pair of qubits is hosted by few-electron-quantum dots under gates P1 and P2. With interstitial J1 controls the exchange coupling between the qubits. On the left is the antenna that delivers high-frequency microwaves, which controls the qubits magnetically through electron-spin resonance (ESR). The single electron transistor (SET) is the charge sensor observing the electron tunneling events. Field programable logical array (FPGA) manipulates the sequential pulsing on gate electrodes and modulation on high-frequency microwaves. (b) Flow chart of the web-based FBT analysis service. The model can be bootstrapped from earlier characterization results. As the posteriors of the last update will become the prior of the next update, the model will be iteratively updated when the measurement outcome arrives. Post-processing on the estimated channels, including CPTP projection and gauge optimization, is performed over the gate set. (c) An example screenshot of the webpage of live update report displayed on a client end. Displays the Hinton diagrams of the estimated channels, error process, and received experiment data. (d) A detailed diagram for channel prior bootstrap and Bayesian update procedures.}
   \label{fbtflow}
 \end{figure*}

\begin{table*}[t]
  \centering
  \begin{tabular}{| c | c | c | c | c|}
  \hline
   \textbf{Strategy}                                                & \textbf{Prior Knowledge}                                                                                                                                                                                                                     &\textbf{Usage Scenario}                                                   &\textbf{Boot time} & \makecell{\textbf{Estimation} \\ \textbf{time}} \\ \hline
  \makecell{ Blind \\ cold boot}                               & \makecell{ $\bar{x}, \Gamma_x$: Guessed channel statistics. \\ General choice will be \\ depolarisation channel}                                                                                                                          & \makecell{No prior knowledge \\ about the new device}     &70 min & 50 min \\\hline
   \makecell{Fidelity constrained \\ cold boot} & \makecell{ $\bar{x}, \Gamma_x$: Guessed channel statistics. \\$\bar{f}, \sigma^2_f$: Gate fidelity statistics. }                    &  \makecell{Managed to run first \\ RBM, based on fidelity \\ numbers} & 50 min & 12 min \\ \hline
\makecell{Partially trusted \\ warm boot}                                         & \makecell{ $\bar{x}$: Mean gate process matrix \\ from finished FBT/GST results. \\ $\Gamma_x$: Guessed uncertainty.  }&  \makecell{Finished some FBT/ \\ GST analysis already.}  & 35 min & 6 min \\ \hline
   \makecell{Fully trusted \\ warm boot}           & \makecell{ $\bar{x}, \Gamma_x$: Gate statistic \\ from finished FBT/GST results. }                                                                       & \makecell{Experiment \\ configuration has not \\ changed  since last \\ FBT run.}  & 0 min & 6 min \\ \hline
  \end{tabular}
  \caption{Comparison of different prior bootstrapping strategies. Conditions for the time consumption estimation: 1. Server computer for analyzing FBT has a 28-core CPU (Intel Xeon W-2175) and 32GB RAM.  2. Time consumptions (wall time) in this table are estimated on FBT analysis on random sequences simulation based on a two-qubit toy model. 3. Boot times depend on the initial guess statistics and the number of samples (typically need 1000 samples) for cold boot and partially trusted warm boot methods. 4. Total estimation time varies, which is strongly dependent on approximation error sampling (here set to 100 samples) and the closeness between the initial guessed model and the true model.} 
  \label{bootcompare}
\end{table*}

\par To minimize the delay between experiment and analysis, we develop a web-based online FBT service, which acts as the infrastructure for the characterization experiments. Details about the FBT protocol can be found in Ref. \cite{timFBT}, with an outline of the protocol and further developments detailed in section \ref{methods}.

\par We call FBT an ``online'' protocol because FBT updates the model as the experimental data becomes available. From an engineering standpoint, deploying the FBT service online provides benefits for the experiment-analysis loop. It frees up the computation load from the experimental setup and makes it possible to run multiple characterization experiments in parallel using an in-house high-performance computer. As shown in Fig. \ref{fbtflow}, the FBT server communicates with the experimental setup and client machines through web application program interfaces (APIs). To start an online FBT analysis session, experimental setup or client machines provide information about the whole gate set, including ideal operators and previous knowledge about the gates for bootstrapping the initial prior for each gate's noise channels. Once the FBT server has finished bootstrapping the initial channels, it is ready for incoming updates from the client machines. FBT updates the model on the fly based on measurement results from the client.  Gate set post-processing includes gauge optimization and complete positive and trace preserving (CPTP) projection, which will only be performed for every $N$ updates (post-processing interval $N$ is customizable) and will not overwrite the original Bayesian statistics. The live updated webpage displays the most recent update of the gate processes and error metrics.

\par As a crucial component of the online FBT service, shown in Fig. \ref{fbtflow}(d), bootstrapping the initial prior sets the starting point for the analysis. In realistic experimental runs, the level of knowledge about the system evolves as more characterization data becomes available. Information obtained from other characterizations, such as Rabi oscillation, randomized benchmarking, and even other process tomography methods, can all serve as prior information for bootstrapping the initial gate prior. The software stack also provides various prior bootstrapping strategies, presented in TABLE. \ref{bootcompare} as a brief guideline for experimentalists. Details about bootstrapping techniques are discussed in section \ref{bootstrap}.

\subsection{Characterizing non-Markovian gate process with FBT}

 \begin{figure*}[tb]
 \includegraphics[width=\columnwidth]{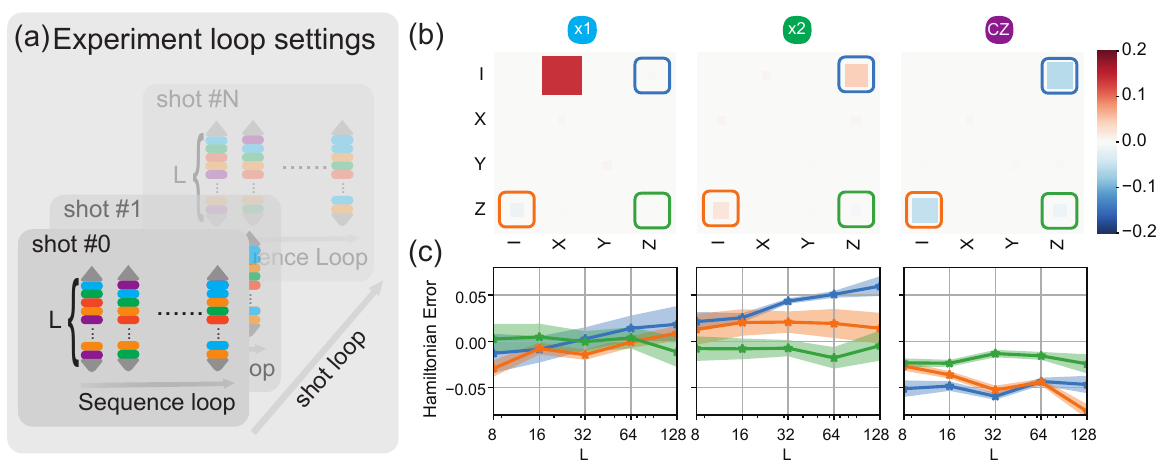}%
 \caption{(a) Illustration of the sequence length dependency experiment. To investigate the dependence of gate errors on the sequence length, we perform five independent experiments, where each experiment has 5000 random gate sequences with the fixed gate length of ${L=8, 16, 32, 64, 128}$. To minimize the impact from slow drifting, the sequences are run in a rasterized manner. (b) Extracted Hamiltonian error from the final estimated results of the experiment $L$=32. (c) Plots of the amplitude of the Hamiltonian error components ($H_{IZ}$, $H_{ZI}$, $H_{ZZ}$) as a function of sequence length. Entries of the error parameter are indicated by their corresponding color by frames on (b). Error bars ($99.7\%$ confidence interval) of the Hamiltonian errors are obtained by resampling the noise channel with the statistic from the last FBT update.}
  \label{lengthDepExperiment}
 \end{figure*}
 
  \begin{figure*}[tb]
 \includegraphics[width=\columnwidth]{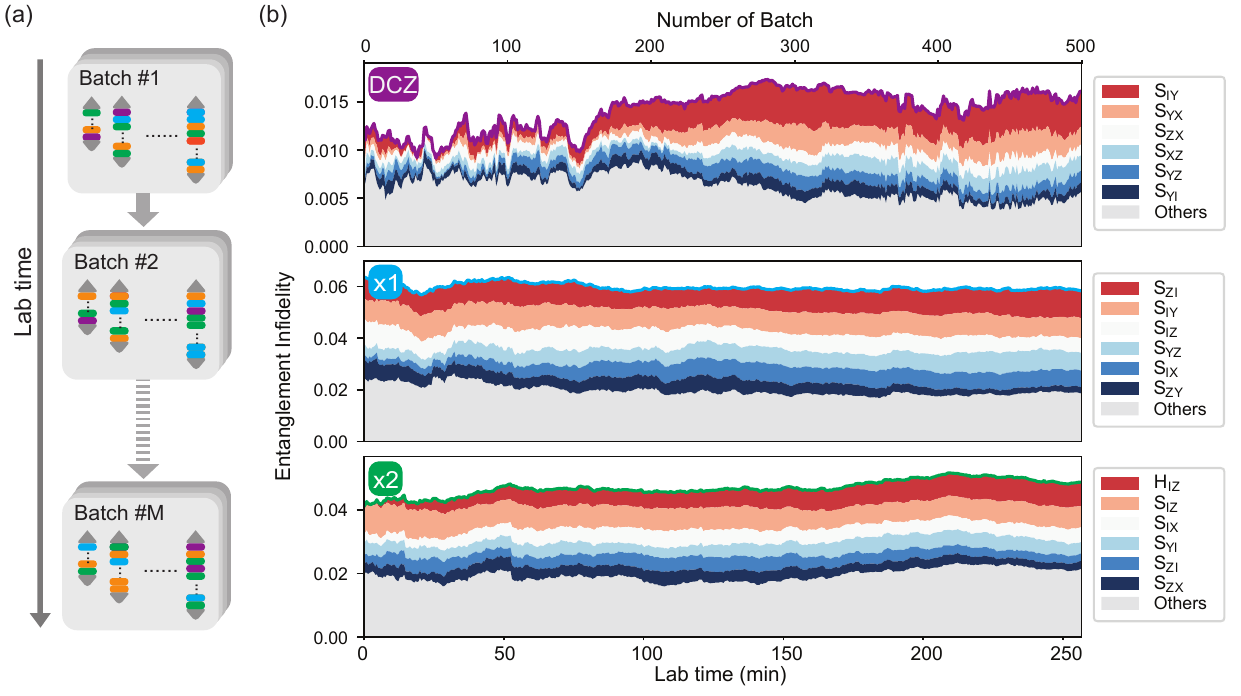}%
 \caption{ (a) Illustration of the slow drift tracking experiment. Sequences with various lengths are executed, batch by batch. Each batch of sequences is looped in a rasterized manner, which runs through 80 sequences for 100 shots. The whole tracking experiment runs over 500 batches of sequences, and each batch is finished in roughly 0.54 minute time window. (b) Entanglement infidelities (colored solid lines on top of each plot) and their top-6 significant error generator infidelity (stacked colored areas under solid line) of gates $x_1, x_2$, DCZ  as a function of lab time of each physical gate. }
  \label{trackingExperiment}
 \end{figure*}
 
 \par Based on our earlier device characterization, we observed that the gates could perform inconsistently depending on the context. Here we study this inconsistency in two different time scales. For the inter-sequence time scale, over an hours-long experiment time, we observe that the gate performance gradually drifts over time. For the intra-sequence time scale, gates perform differently in sequences with different lengths. This behavior was observed on silicon spin qubits from RB experiments on a single qubit \cite{fogarty2015nonexponential}. In that work, the presence of the non-exponential RB decays was attributed to be an indication of the impact of non-Markovian errors when running long sequences. 
 
\par In the presence of non-Markovian noise, process tomography protocols find it challenging to reconstruct such context-dependent processes. Though standard process tomography protocols do not capture the non-Markovian dynamics, one might hope to learn how badly the estimation is being affected. GST \cite{GST0, GST1} provides model violation as the metric of the goodness of fit. A large model violation in a data set is one indicator that the underlying process could be highly non-Markovian. However, this metric does not provide any information about what types of errors appear non-Markovian. One feature of FBT is that it can take arbitrary sequences as input to update its model. This freedom in the choice of sequences allows us to design experiments to uncover gate error inconsistencies under different contexts. We focus on the intra-sequence and inter-sequence regimes individually and investigate the non-Markovian nature of the gate processes.  For the intra-sequence regime, we design a sequence-length-dependent experiment to probe how the error changes as we run different lengths of sequences. For the inter-sequence regime, we track the slow drifting of the gate performances by warm booting the model and running experiments in batches within a fixed time window.

\subsubsection{Sequence length dependency}
\label{lengthDepErr}
\par To look into the sequence length dependency of gate errors, we run a series of experiments with different fixed sequence lengths (length in the number of primitive gates) and analyze each individually. The fixed sequence lengths are chosen to be 8, 16, 32, 64, and 128. Each experiment consists of 5000 randomly generated sequences with the same length, generally enough to ensure informational completeness data for two-qubit system's tomographic characterization. Each sequence is repeated for 100 shots, and no extra projections are needed (see section \ref{primitiveReadout}). Though 5000 sequences with 100 shots only takes around 20 minutes to acquire, the device is still impacted by undesired slow sub-Hertz noise. There will be significant inter-sequence inconsistency if the shot repeating loop is nested inside the sequence selection loop, for example, if we were to acquire all repeated shots for one sequence and then loop to the next sequence. To minimize the inter-sequence inconsistency for this experiment, we repeat the list of sequences in a rasterized manner,  also used in Ref. \cite{proctor2020detecting}. Sequence rasterization means all sequences are looped for one single shot, then the next shot repeats all sequences. This can be implemented on FPGA by programming the sequence compilation loop to be embedded in the shot-repeat loop. 

\par We choose the last update for each experiment to investigate the sequence length dependent errors. FBT is primitively estimating the noise residual PTM (Pauli transfer matrix) for non-ideal gate processes, which describes the input-output relationships on Pauli basis. To make it more physically intuitive, we convert the final estimated gate noise channel to elementary error generators (see section \ref{errorTaxonomy}). Since we find that the Hamiltonian error captures $77\%$ to $97\%$ of the error generator, we mainly focus on the Hamiltonian errors. As shown in Fig. \ref{lengthDepExperiment}(b), the dominating error is $H_{IX}$ error on $x_1$ (see Methods. \ref{gatesetDefine} for gate definitions), which indicates an over-rotation error. Smaller errors, including $H_{IZ}, H_{ZI}, H_{ZZ}$ errors, appear to have sequence length dependency. Noticeably,  we find the $H_{IZ}$ error on $x_2$ gate has a strong sequence length dependency, which likely originated from the Larmor frequency shift induced by transient microwave effects \cite{UndsethNonlinear, takedaFreqShift, SolomonHeating, fogarty2015nonexponential}.

\subsubsection{Slow drift tracking}
\label{slowDrift}
\par The slow-drift tracking experiment aims to observe sub-Hertz noise, which manifests as gate performance metrics drifting over hours-long experiments. The validity of this experiment-analysis is built on the assumption that the system is relatively stable within a few-minutes-long time window, and drifts so slowly that it allows the changes of the error parameters to be smooth and continuous. Within each "stable" window, we sample as much data as possible. We rule out the length dependency errors by running various lengths of random sequences. With the arguments in hand, we use the experiment design as shown in Fig. \ref{trackingExperiment}(a). Random sequence batches, which consist of various lengths of random sequences, are streamed into an FPGA after the previous batch is finished. Each batch of sequences was executed in a rasterized manner as in section \ref{lengthDepErr}.

\par To track the drift of the error parameters from the beginning, the FBT prior was bootstrapped with an educated guess that is closest to the first batch of experiment. The best practice for this is a warm boot. The details about the warm boot strategy are discussed in section \ref{bootstrap}. Since warm boot already sets the model with the best prior information, the estimated results of the first batch of experimental data can be approximately treated as the pseudo-transient state at its corresponding lab time. The following batches are analyzed in lab time order so that we can correlate the batch analysis with lab time.

\par Fig. \ref{trackingExperiment}(b) shows gate entanglement infidelities \cite{entanglementFidelity} and their generator infidelities as a function of lab time. Stochastic errors contribute most to gate infidelity for all three gates. $S_{ZI}$ type error on $x_1$ and $S_{IZ}$ error on $x_2$ are due to the idling on the spectator qubit when operating the single qubit gates. The Hamiltonian error on $H_{IZ}$ on $x_2$ gate could have a similar physical origin as we find in section \ref{lengthDepErr}. The stochastic errors on DCZ vary quite significantly, possibly due to fridge temperature fluctuations or charge movement in silicon. 

\section{Methods}
\label{methods}
\subsection{Device and gate implementation}
\label{gatesetDefine}
\par  Shown in Fig. \ref{fbtflow}(a) (same as device A and B in \cite{tanttu2023stability}), the devices (A and B) we use are Silicon-Metal-Oxide-Semiconductor (SiMOS) devices with a purified $^{28}$Si substrate and aluminum gate electrodes. Few-electron double quantum dots are formed at the interface of Si/SiO$_2$ under gates P1 and P2. An external static magnetic field is applied in-plane and creates around 20 $\mathrm{GHz}$ Zeeman splittings. Two spins can be addressed individually due to the Zeeman energy difference (around 10 $\mathrm{MHz}$ for A, and 22 $\mathrm{MHz}$ for B). The exchange coupling between the two spins is controlled by gate J1.  The spin readout is achieved by tunneling one electron from one dot to another, where the tunneling event is observed by a single-electron-transistor (SET) when the two spins are in odd parity \cite{seedhouse2020parity, leon2021bell}.
 
\par In this work, the gate set includes five elementary gates, which are $X^{\pi/2}_{Q1}$, $X^{\pi/2}_{Q2}$, $Z^{\pi/2}_{Q1}$, $Z^{\pi/2}_{Q2}$ and controlled phase gate CZ (in device A) or decoupled-controlled phase gate DCZ  (in device B) \cite{watsonDCZ, Vandersypen6QHeatingEff}. For clarity, we label them as $ \{x_1, x_2, z_1, z_2,$ CZ (or DCZ)$\}$ individually. Here $x_1$ and $x_2$ are implemented by electron-spin resonance by pulsing modulated oscillating magnetic field, which is delivered by the ESR antenna. Gates $z_1$ and $z_2$ are software-implemented virtual gates,  which are implemented by changing the rotating frame of each resonator that tracks qubit frequency \cite{RB2QWister}. Both the CZ and DCZ gates are implemented by pulsing the J1 gate, with the key difference that the DCZ has a $\pi$-pulse on both qubits sandwiched in the middle of two J1 pulses. Pulsing on J1 introduces an extra Stark shift on each of the qubits, which are compensated by software phase correction for the CZ gate, while this will be canceled out with the DCZ gate. Residual $^{29}$Si nuclear spin flips result in jumps of spin resonance frequency, and we have implemented feedback protocols to track and correct the Larmor frequencies. In addition, we have feedback on Rabi frequencies and exchange coupling. Feedback protocols are interleaved between the main experiment runs.

\subsection{Fast Bayesian tomography protocol}
\label{FBToutline} 
\par Before introducing the new improvements on FBT, in this subsection, we review the basics of the protocol \cite{timFBT}. FBT is a self-consistent method that uses Bayesian inference to reconstruct the whole gate set simultaneously. A gate set is a full mathematical description of the capability of a quantum system, including state initializations, gate operations, and measurements. Generally, an experiment begins with initialization of the quantum system to state $\rho$, then follows a sequence $S_k$ composed of the unitary gates and, finally, measurement is a projection to effect $E$. Collecting the counts of the states yields the probability of the specified outcome of that experiment. This is described by:
\begin{equation}
m_k =  \bra{\bra{E}}  \prod_{i\in S_k}  G_i \ket{\ket{\rho}}
\label{selfconsistModel1}
\end{equation}
where $m_k$ is the outcome probability, $\bra{\bra{E}}$ and $\ket{\ket{\rho}}$ are vectorized measurement operator and preparation density matrix and $\prod_{i\in S_k}$ represents the sequential operation of the gates in sequence $S_k$.

\par To account for errors in the model, a noisy gate $\tilde{G_i}$ is modeled as an ideal gate $G_i$ followed by a noise channel  $\Lambda_i$. Since the SPAM processes ($\bra{\bra{\tilde{E}}}$ and $\ket{\ket{\tilde{\rho}}}$) are imperfect, we also associate them with noise channels to capture the errors, represented by $\Lambda_{E}$ and $\Lambda_{\rho}$ respectively. So the noisy gate set is described by:
\begin{eqnarray}
	 \left\{
	  \begin{aligned}
	  		\bra{\bra{\tilde{E}}} = \bra{\bra{E}} \Lambda_E  \\
			\tilde{G}_i = G_i \Lambda_i \\
			\ket{\ket{\tilde{\rho}}} = \Lambda_{\rho} \ket{\ket{\rho}}
  	\end{aligned}
  \right.
\end{eqnarray}
% some space is missing after this equation
\newline
\par The tomographic reconstruction of the noise channels requires solving the non-linear problem, which is computationally challenging. This problem can be solved by linearizing Eq.\ref{selfconsistModel1}, which decomposes the noise channel into $\Lambda_i = I + \varepsilon_i$ and drops higher-order terms \cite{MerkelSelfConsistent}. For a sequence with a length of $N_k$, we have 
\begin{multline}
\label{selfconsistModel2}
m_k \approx  \bra{\bra{E}}  \prod_{s \in S_k}  G_s \ket{\ket{\rho}} 
\\ + \bra{\bra{E}} \varepsilon_E \prod_{s \in S_k}  G_s \ket{\ket{\rho}} 
\\ + \sum_{j = 1}^{N_k} \bra{\bra{E}} \left[ \prod_{i = j +1}^{N_k} G_i \right]  \varepsilon_j G_j  \left[ \prod_{i =1}^{j-1}  G_i \right] \ket{\ket{\rho}}  
\\ + \bra{\bra{E}} \prod_{s \in S_k}  G_s  \varepsilon_{\rho} \ket{\ket{\rho}} 
\\ = m_{\text{ideal}} + A_k x
\end{multline}
where $m_{\text{ideal}}$ is the ideal output and $x$ is the  vectorised form of noise channel residual  $\varepsilon_i$. However, this model only works for gates with high fidelity and fails when errors are large.

\par Based on the linearized model, we construct a Bayesian model to estimate the noise channel parameters. Firstly, let each noise channel residual parameter be distributed as a Gaussian variable, then $x$ will become a multi-variable Gaussian $\ket{\ket{\varepsilon_i}} \sim \mathcal{N}(\ket{\ket{\bar{\varepsilon}_i}}, \Gamma_{i})$. If we are bootstrapping the multi-variable Gaussian from an educated guessed prior, then the FBT model is:

\begin{widetext}
	\begin{multline}
m_k \approx  \bra{\bra{E}}  \bar{\Lambda}_E \prod_{s \in S_k} \bar{\Lambda}_s  G_s \bar{\Lambda}_{\rho} \ket{\ket{\rho}} + 
 \bra{\bra{E}} \varepsilon_E \prod_{s \in S_k} \bar{\Lambda}_s  G_s \bar{\Lambda}_{\rho} \ket{\ket{\rho}} + 
\\ \sum_{j = 1}^{N_k} \bra{\bra{E}} \bar{\Lambda}_E \left[ \prod_{i = j +1}^{N_k} \bar{\Lambda}_i G_i \right]  \varepsilon_j G_j  \left[ \prod_{i =1}^{j-1}   \bar{\Lambda}_i G_i \right] \bar{\Lambda}_{\rho} \ket{\ket{\rho}}
 + \bra{\bra{E}} \bar{\Lambda}_E \prod_{s \in S_k} \bar{\Lambda}_s  G_s  \varepsilon_{\rho}  \ket{\ket{\rho}} 
\\ = \bar{m}_k + \bar{A}_k x + \epsilon_k + \eta_k
\end{multline}
\end{widetext}

\par Based on this, the FBT protocol updates itself iteratively. For example, by forwarding the posterior of the $(k-1)^{th}$ update as the prior for the new coming $k^{th}$ update, the model infers the expected outcome $\bar{m}_k$ of $k^{th}$ sequence and the linear model $\bar{A}_k$ which acts on the centralized residual error parameters $x$. Besides gate set errors, FBT also models two error processes individually, which are approximation error and sampling error.

\par We note that, the approximation error captures the error due to model linearization, which avoids overfitting issues when gate errors are relatively large. Since estimating the approximation error requires sampling over the estimated channels, which is time-consuming, it becomes unnecessary and can be dropped off when it becomes much smaller than shot noise \cite{timFBT}.

\subsection{Decomposing into elementary error generators}
\label{errorTaxonomy}
\par FBT represents noise channels in PTM, which makes channel parameterization easier. However, it is not intuitive to understand the physical sources of the errors. Error taxonomy method for small Markovian errors \cite{errorTaxonomy} decomposes the error generators into elementary error generators, which allows us to correlate the physical error sources from the estimated noise channels.

\par To decompose the noise channels estimated by FBT,  PTM of noise channels  $\Lambda$ need to be converted to error generator $\mathcal{L}$ by matrix logarithm:

\begin{equation}
	\mathcal{L} = \text{logm}(\Lambda)
\end{equation}

Error generators can be projected to error subspaces. The elementary error generator can be categorized into four classes: Hamiltonian, Pauli-stochastic, Pauli-correlation, and active error generators (denoted as H, S, C, and A individually). The coefficients of elementary error generators can be extracted by its dual basis. For instance, for Hamiltonian error generators:
\begin{equation}
	H'_P[\mathord{\cdot}] = -\frac{i}{2 d^2}[P, \mathord{\cdot}]
\end{equation}
where $P$ denotes Pauli operators and $d$ is the dimension of Hilbert space. 
 
 \par Then the coefficient of the elementary Hamiltonian error generator can be calculated by:
 \begin{equation}
 	h_P = Tr(H_P^{\dagger} \mathcal{L})
 \end{equation}
 where $H_P^{\dagger}$ is the dual Hamiltonian generator in Pauli basis. 
 
 \par We can also evaluate the elementary error generator infidelity contributions to entanglement infidelity by:
 \begin{equation}
 	\epsilon_{\text{ent}} = \epsilon_{J} + \theta_J^2 + O(|L_{S}|^2)
 \end{equation}
 where the Jamiolkowski probability $\epsilon_{J}$ and Jamiolkowski amplitude $\theta_J^2$ are:

\begin{equation}
	\epsilon_{J} = \Sigma_P S_P
\end{equation}

\begin{equation}
	\theta_J^2= \Sigma_P H_P^2
\end{equation}

\par The error taxonomy as a post-processing procedure for FBT allows us to reinterpret the outcomes in a more sensible way.  The coefficients of elementary error generators tell what are the dominating errors that limit us from higher fidelity and indicate the potential physical origins of the errors.

\subsection{Informational completeness}
\label{sec: informationComplete}
\par Treating tomography protocols as black boxes, simply intaking measurement results and outputting the reconstructed mathematical description of the model, often raises the question of how much data is required to reconstruct the model faithfully. The minimal information required by traditional state, process, and measurement tomography should span the whole Hilbert-Schmidt space. For example, reconstruction of an $n$-qubit state requires a minimum of $4^n$  projective measurements that are orthogonal to each other.

\par Under the picture of self-consistent tomographies, like FBT, the requirements for informational completeness is no longer as straightforward as process tomography based on linear inversion reconstruction. We shall see that each measurement outcome is a collective effect of a subset of unknown parameters from the gates used in that sequence. Since FBT is a Bayesian method, we can update our estimates without requiring informational completeness. Though quantitively determining the minimum number of experiments for FBT is beyond the scope of this work, we list the following empirical principles:

\begin{itemize}
\item Distance between the initial guess and the true model: large distances require more updates to make the final estimation reach a decent level of accuracy.
\item The initial guess of the uncertainty in the model parameters needs to balance the speed of the convergence towards a result while being loose enough to allow for significant updates on the model with incoming new data.
\item Randomness of the sequences set: unbiased appearances of each gate improves the chances of all parameters reaching the target level of precision with a finite number of experiments.
\end{itemize}

\par Eventually, the precision will be limited by finite sampling and the non-Markovianity of the system. In this work, we are chasing to make the cost of experiments lower without compromising the estimation accuracy with the advantages of FBT. The next section approaches this target by reducing unnecessary measurement projections, while Section \ref{bootstrap} discusses how the characterization information can be maximally used by the initial prior bootstrap.

\subsection{Native measurement for FBT}\label{primitiveReadout}

 \begin{figure*}[tb]
 \includegraphics[width=\columnwidth]{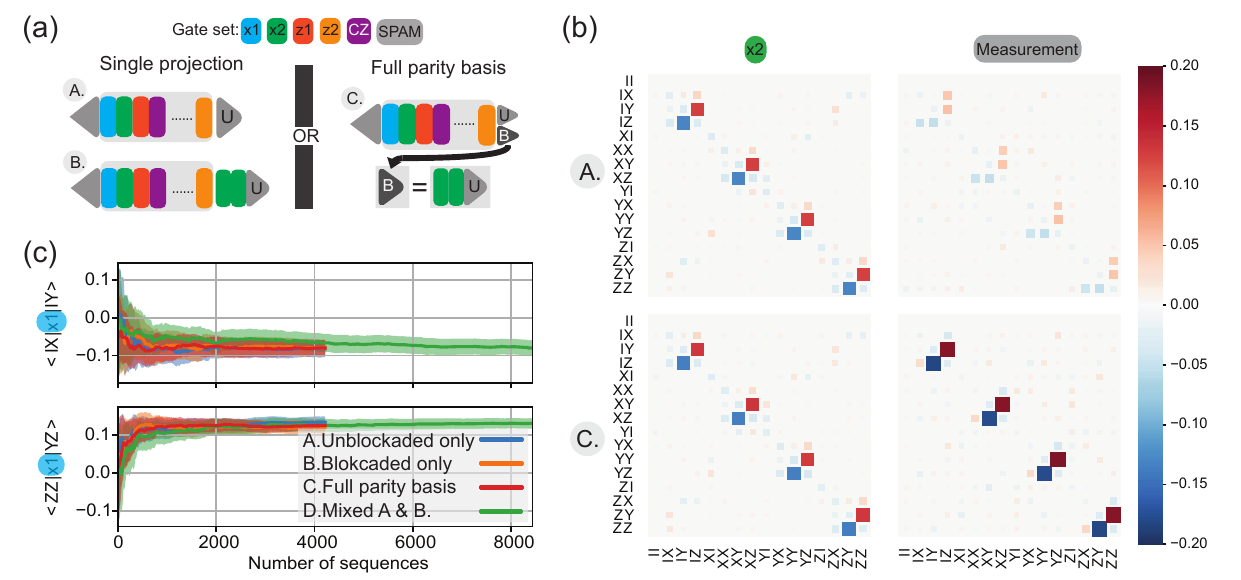}%
 \caption{Validity of native measurement for FBT. (a) Illustration of three ways of feeding projective measurement outcomes to FBT. Each main sequence is repeated twice, one for measuring odd parity and another for even parity. Cases A and B take one of the two projections for FBT analysis. Case B absorbs the projection sequence into the main sequence. While case C takes both of the two projections as a full parity basis. (b) Final estimated gate noise residual channel of $x_2$ and measurement channel for cases A and C. (c) Traces of channel parameters of $x_1$ gate for each case mentioned in (a). Case D mixes data from cases A and B as a reference, which processes twice the amount of the sequences.}
 %is this last sentence a comment?
  \label{readout}
 \end{figure*}
 
 \par Most process tomography protocols assume that each qubit can be measured individually, which means each qubit gives 1 bit of classical information for each shot. However, for a multi-qubit system, it is more straightforward to measure the qubits pairwise. Parity readout \cite{seedhouse2020parity} based on the Pauli exclusion principle yields one bit of classical binary information to determine whether a pair of spins are parallel or antiparallel. Earlier work  \cite{leon2021bell} has shown that we can access the complete two-qubit measurement basis by repeating the main sequence with different projection sequences. For QPT, it is necessary to have an informationally complete measurement to guarantee linear inversion. However, for self-consistent methods like FBT, as we have discussed in section \ref{sec: informationComplete}, it is not necessary to measure multiple projections for each random circuit to satisfy information completeness. 

\def\mvec{
\begin{bmatrix}
	m_{k0} \\ m_{k1} \\ \cdots \\ m_{kN_E}
\end{bmatrix}}
\def\Evec{
\begin{bmatrix}
	\bra{\bra{E_0}} \\ \bra{\bra{E_1}} \\ \cdots \\ \bra{\bra{E_N}}
\end{bmatrix}}

\par We show here that the measurement for FBT can be formulated either with multiple projections or directly using a single native measurement outcome. To clarify the problem, we naively look into the problem with Eq. \ref{selfconsistModel1}. A sequence repeated with $N_E$ projections can be written as:
\begin{multline}
\mvec = \Evec \Lambda_{E} \times \\
\prod_{i\in S_k} \Lambda_i G_i \Lambda_{\rho}\ket{\ket{\rho_0}}
\end{multline}
As indicated by the model, every shot of the experiment starts with state initialization, executing the main sequence, and ending with a measurement. Suppose here we have only one native measurement, the rest can be implemented by performing projection sequences before the native measurement. Alternatively, the equation above can be written as:
\def\Eprojvec{
\begin{bmatrix}
	\Lambda_{E} I \\ \Lambda_{E} \prod_{i\in P_1} \Lambda_i G_i  \\ \cdots \\ \Lambda_{E} \prod_{i\in P_N} \Lambda_i G_i 
\end{bmatrix}}

\begin{multline}
   \mvec = \bra{\bra{E_0}} \Eprojvec \times \\ 
   \newline
   \prod_{i\in S_k} \Lambda_i G_i \Lambda_{\rho}\ket{\ket{\rho_0}}
\end{multline}

\par By joining each projection sequence to the main sequences individually, the original M sequences dataset is now unpacked to be $N_E \times M$ sequences with native measurement. To prove that feeding FBT with a single projection does not harm informational completeness, we compare three ways of using the projective measurement results as the input to FBT.

\begin{itemize}
\item 1. FBT receives $M$ updates with original main sequences, each update takes multiple projections as a vector input.
\item 2. FBT receives $M$ updates, keep one of the projections and join that one projection sequence to the main sequences, each update takes that projection's result as a scalar input.
\item 3. FBT receives $N_E \times M$ updates, utilize all projection's results and inputs them to FBT like 2. 
\end{itemize}

\par Without losing generality, we demonstrate with parity readout, which is used in all the experiments shown in the previous sections. For instance, for a pair of electron spins, if tunneling of one electron from one dot to its neighbor happens, this indicates an odd parity state. To measure the probability of the opposite parity, we invert the parity by applying a $\pi$ pulse on one of the qubits, which is implemented by a projection sequence of $x_2-x_2$.  For a parity readout natively reading out odd parity, the projections can be represented as: 

\begin{equation}
	\begin{split}
	\bra{\bra{E_{Odd}}} = \frac{1}{2} ( \bra{\bra{\uparrow\downarrow}} + \bra{\bra{\downarrow\uparrow}}) \\
	\bra{\bra{E_{Even}}} =  \frac{1}{2}( \bra{\bra{\uparrow\uparrow}} + \bra{\bra{\downarrow\downarrow}}) = \\ 
	 \frac{1}{2} ( \bra{\bra{\uparrow\downarrow}} + \bra{\bra{\downarrow\uparrow}}) (G_{x_2} G_{x_2})
	\end{split}
\end{equation}

\par As shown in Fig. \ref{readout}, the testing experiment contains 4220 random sequences as main sequences, and each sequence is repeated twice to get both even and odd parity projections, which means 8440 different sequences in total were executed for this experiment. Cases A and B each use one of the projections individually. However, for case B, the projection sequence is seen as part of the main sequences, so FBT intakes a single native measurement outcome like case A.  While case C routinely takes two projections as a complete parity measurement basis, case D is a reference case, which utilizes all 8440 sequences, but FBT sees the projection sequences as part of main sequences, like cases A and B. 

\par Based on the four cases of analysis of the testing experiment, the estimated parameters' accuracy are not sacrificed even with the single-projection native measurement cases (comparing A and B to C), which indicates that multi-projection parity measurement does not benefit the accuracy of estimation. 

\par From Fig. \ref{readout}(b), it is also noticeable that both cases A and C show that $x_2$ has quite a significant over-rotation error. However, for case C, the $x_2$ over-rotation error appears on the measurement channel, which is not desirable when trying to diagnose the intrinsic errors in the measurement channel. 

\par Therefore, a single native measurement as input for FBT does not harm the accuracy of the estimation. By dropping off the non-native measurements, the experiment and analysis costs were reduced to $1/N_E$, and mingling of gate error to measurement channel can be avoided. 

\subsection{Gauge optimization for FBT}

\par Under the self-consistent picture, the representation of the reconstructed gate set is not unique. That means multiple alternative representations of the gate set yield the same experimental outcomes. The transformation between those equivalent representations is called gauge transformation:
\begin{eqnarray}
	 \left\{
	  \begin{aligned}
	  		\bra{\bra{\tilde{E}_0’}} = \bra{\bra{\tilde{E}_0}} S  \\
			\tilde{G}_i’ = S^{-1} \tilde{G}_i S \\
			\ket{\ket{\tilde{\rho}’_0}} = S^{-1} \ket{\ket{\tilde{\rho}_0}}
  	\end{aligned}
  \right.
\end{eqnarray}
where the transformation matrix $S$ is arbitrary as long as $S$ is invertible and trace-preserving, which maintains the CPTP of the transformed gate set. 

\par In principle, there is no preferred choice of gauge for a gate set. However, metrics that indicate the overall performance of the gates, like fidelity, are gauge-variant. This is particularly impactful in the case where the state preparation and measurement are being studied as noisy channels themselves, such as we do here. In this case, we are prescribing the initialization and measurement quantization axes. We resolve the issue of gauge ambiguity for our FBT protocol by fixing the gauge to be the one that optimizes some choice of metric for gate fidelity.

\par In this work, we choose the target of the gauge optimization to be minimizing a ``distance'' metric between the transformed gate set and the ideal gate set. Though there are a few metrics, like fidelity, diamond norm, etc., that can be chosen as the objective function, minimizing the weighted Frobenius distance between the estimated gate set and ideal gate set would be the optimal option \cite{GST1}:
	
\begin{equation}
	\begin{aligned}
	\label{gaugeoptfunc}
	\operatorname*{argmin}_S g(\mathcal{G}, \mathcal{G'}) = w_G \sum_{i} ||\tilde{G}_i - \tilde{G}_i'||^2 _{\mathcal{F}} \\ + w_{S} ( ||\tilde{\rho}- \tilde{\rho}'||^2_{\mathcal{F}} +  ||\tilde{E}- \tilde{E}'||^2_{\mathcal{F}} )
	\end{aligned}
\end{equation}
where $|| \cdot ||_{\mathcal{F}}$ denotes Frobenius norm, $w_G$ and $w_S$ are unitary gate weight and spam weight individually. Typically we set $w_G/w_S >> 1$, because the SPAM errors can not be amplified. Similar to the implementation in pyGSTi \cite{pyGSTi}, gauge transformation matrix $S$ is constrained to be trace-preserving and invertible. The weighted sum of the Frobenius distance, as the object function, is minimized by the L-BFGS-B method \cite{LBFGSB} from the $Scipy$ python package. 

\subsection{Initial prior bootstrap strategies}

\label{bootstrap}

\par One major advantage of the Bayesian method is that we can incorporate as much knowledge as we have known into the model prior before we start the analysis. Every coin has two sides, the bootstrapping strategy for the initial prior of the channels is critical \textemdash a good educated guess can help get trustworthy results with low experimental and computational cost, while a poor guess could lead to convergency problems. 

\par There are several reasons why a good initial guess is important for FBT. Firstly, we can feed FBT with ``more than enough'' amounts of updates to guarantee the accuracy of estimation, but they would also take a long time to run experimentally, which in turn leads to slow drifts becoming significant. Since a lesser amount of experiments is always more desirable, an educated guess for the initial prior allows the model to be updated with less data, but without compromising the accuracy. Secondly, a decent initial guess allows the model to start with lower approximation error. The approximation error, which captures the error from linearizing the model, is expensive to sample for each update and can not be dropped before meeting the turn-off threshold \cite{timFBT}. Thus, a wise choice of the bootstrap strategy, as briefly summarised in Table \ref{bootcompare} is critical for FBT.

\par In practice, the quantum system we are characterizing is not entirely unknown to us. As more characterization data becomes available, we have more trustable information to bootstrap the initial prior statistics for FBT. In the worst case, we don't have any useful information about the whole gate set and have to cold boot blindly with roughly estimated noise channels based on metrics like quality factors and visibility of Rabi oscillations, where depolarising noise channels is usually the conventional choice. In this case, we do have to feed more sequences to FBT to get reliable results. Another cold boot strategy \textemdash cold boot with fidelity \cite{timFBT} estimated by earlier RBM experiments \textemdash provides a tighter error bound than a blind cold boot. This significantly reduces the approximation error for beginning updates and doesn't require a large number of sequences to feed FBT. 

\par Routine calibration and diagnosis with FBT will accumulate abundant results in the database, which is also a source of knowledge for bootstrapping the prior. In this work, we introduce warm boot strategies to leverage prior information from historical process tomography results. We can either trust the earlier analysis fully (full warm boot) or partially (partial warm boot), depending on the device setup. The fully warm boot is applicable when no significant changes happen to the system configurations. The new analysis fully inherits the complete gate set statistic of the previous analysis. Minor updates of the setup or slow drift can impact the gate performance locally but we can still partially trust the earlier results by overwriting a different uncertainty based on educational guesses. Algorithm \ref{warmbootalg} shows how partial warm boot initializes the initial prior with the previous estimation results and new guess of the covariance matrix. 

\begin{algorithm}
  \DontPrintSemicolon
  \SetAlgoLined
  	\KwIn{ $\bar{x}$: Estimated noise channel mean from previous analysis}
  	\KwIn{ $\Gamma_x$: Guessed covariance matrix}
	$x \sim \mathcal{N}(\bar{x}, \Gamma_x)$ \\
	\For{$i = 0$ \KwTo $N_{sample}$}{
		Gaussian sample a process matrix $\mathcal{X}$ \\
		CPTP project $\mathcal{X}$ \\
		Save $\mathcal{X}$ to $P_f(\mathcal{X})$
	}
	$\bar{x'} \leftarrow mean(P_f(\mathcal{X}))$ \\
	$\Gamma_x' \leftarrow cov(P_f(\mathcal{X}))$ \\
    \KwOut{$(\bar{x'} , \Gamma_x')$}
  \caption{Initial prior bootstrap: partial warm boot}
  \label{warmbootalg}
\end{algorithm}

\section{Discussion}

\par The web-based online analysis platform developed in this work opens up the possibility for real-time gate set calibration in the future. We characterize the non-Markovian quantum process by two specially designed FBT experiments in different scales of noise effective time on a pair of spin qubits in a silicon quantum dot system. We observed that the Hamiltonian error dominates, and some of its components appear stronger in longer sequences. Slow drift tracking experiment shows that qubit fidelity can vary over a long experiment time, which can be captured by FBT. To reduce the experimental cost and acquire more information per unit of time, we verified that the native measurement method does not compromise the estimation accuracy and proposed the warm boot for the initial prior bootstrap to speed up the analysis.
 
\par Though our experiment-analysis protocol for observing non-Markovian error is not a rigorous tomographic method for reconstructing non-Markovian quantum process dynamics, it clearly indicates the types of errors that have non-Markovian behavior. To close the loop between characterization and correction, we will need to correlate the estimated errors back to instrumental control parameters in future work. Scaling up the quantum system will also bring challenges to FBT. The real-time online feedback may no longer be valid as the analysis time scales non-linearly with the number of qubits. Refining the model and reducing the number of parameters is the potential solution to this challenge. Overall, in this work, we have demonstrated the potential of FBT for tackling non-Markovian noise and made it ready to be an online real-time feedback tool for building future fault-tolerant quantum computers.

\section{Acknowledgement}
\par We acknowledge technical discussions with Matthew Otten from HRL Laboratories. We acknowledge support from the Sydney Quantum Academy, the Australian Research Council (FL190100167, CE170100012, and IM230100396), the US Army Research Office (W911NF-23-10092), and the NSW Node of the Australian National Fabrication Facility. The views and conclusions contained in this document are those of the authors and should not be interpreted as representing the official policies, expressed or implied, of the Army Research Office or the US Government. The US Government is authorized to reproduce and distribute reprints for Government purposes notwithstanding any copyright notation herein.

\bibliography{mybibliography}{}
\bibliographystyle{unsrtnat}
\end{document}